\documentclass{article}

\usepackage{arxiv}

\usepackage[utf8]{inputenc} 
\usepackage[T1]{fontenc}    
\usepackage{hyperref}       
\usepackage{url}            
\usepackage{booktabs}       
\usepackage{amsfonts}       
\usepackage{nicefrac}       
\usepackage{microtype}      
\usepackage{lipsum}
\usepackage{graphicx}
\graphicspath{ {./images/} }

\title{Explaining the (not so) obvious: Simple and fast explanation of STAN, a next Point of Interest recommendation system}

\author{
 Fajrian Yunus \\
  LTCI, T\'el\'ecom Paris\\
  Institut Polytechnique de Paris\\
  Palaiseau, \^{I}le-de-France, France \\
  \texttt{fajrian.yunus@telecom-paris.fr} \\
   \And
 Talel Abdessalem \\
  LTCI, T\'el\'ecom Paris\\
  Institut Polytechnique de Paris\\
  Palaiseau, \^{I}le-de-France, France \\
  \texttt{talel.abdessalem@telecom-paris.fr} \\
}

\begin{document}
\maketitle
\begin{abstract}
A lot of effort in recent years have been expended to explain machine learning systems. However, some machine learning methods are inherently explainable, and thus are not completely black box. This enables the developers to make sense of the output without a developing a complex and expensive explainability technique. Besides that, explainability should be tailored to suit the context of the problem. In a recommendation system which relies on collaborative filtering, the recommendation is based on the behaviors of similar users, therefore the explanation should tell which other users are similar to the current user. Similarly, if the recommendation system is based on sequence prediction, the explanation should also tell which input timesteps are the most influential. We demonstrate this philosophy/paradigm in STAN (Spatio-Temporal Attention Network for Next Location Recommendation), a next Point of Interest recommendation system based on collaborative filtering and sequence prediction. We also show that the explanation helps to ``debug'' the output.
\end{abstract}


\section{Introduction}
Machine learning has seen widespread adoption for the past many years. However, unlike the classic linear regression, it is often hard to understand the internal working of modern machine learning techniques like deep learning. This problem led to the emergence of explainable machine learning as a research area. There are techniques such as Shapley\textquotesingle s values~\cite{strumbelj2010efficient} or LIME~\cite{ribeiro2016should} which explain a black box machine learning method. However, such techniques can be expensive. For example, the calculation of Shapley\textquotesingle s values often involve sampling. Besides that, the explanation should to take into account the context of the problem itself. In collaborative filtering methods, the information about which other users are similar to the current user is important. Similarly, if the problem is modeled as a sequence prediction problem, the information about which input timesteps are influential is an important information.

Spatio-Temporal Attention Network for Next Location Recommendation (STAN)~\cite{luo2021stan} is a next Point of Interest (POI) recommendation system based on deep learning. STAN gives a personalized recommendation of the next POI based on the user\textquotesingle s past visited POIs and the visit timestamps. The number of the past visits varies between users. STAN internally uses attention mechanism~\cite{bahdanau2015neural}. Attention mechanism has a side effect that it ``explains'' which input timesteps are the most important for the output. STAN also has an embedding module whose output represents the ``behavior'' of the user. This creates a notion of similarity between the different users, and thus implements collaborative filtering where the behaviors of similar users is taken into account to decide the recommendation.

We perform a local explanation based on the relevant other users and the relevant timesteps of the current user\textquotesingle s past trajectory (i.e., ``by relevant user or item'' according to \cite{zhang2020explainable}\textquotesingle s classification) . We use the standard dataset of STAN. This dataset contains only the POIs\textquotesingle latitude and longitude and the users\textquotesingle s trajectory. The trajectory is modeled as a sequence of POI ID and timestamp. Therefore, the explanations are which other users are similar to the current user, and which timesteps of the current user\textquotesingle s past trajectory are influential. Because what we know about the users are only their trajectories, intuitively similar users should also have similar trajectories.

In this paper, we use STAN\textquotesingle s attention block to identify the important timesteps of the user\textquotesingle s past trajectory. We also identify the most similar users (relative to the current user) by using STAN\textquotesingle s user embedding module. Finally, we verify these results experimentally. Our code is available on \url{https://bitbucket.org/fajrianyunustelecomparis/stan-explainable/}.

\section{Background} \label{sec:background}
STAN is a next Point of Interest (POI) recommendation system. Its output is the recommended POI\textquotesingle ID while the inputs are the user\textquotesingle s ID and the user\textquotesingle s past trajectory. The trajectory is a variable length sequence where each element is the visited POI\textquotesingle s ID and the visit\textquotesingle s timestamp. The timestamp\textquotesingle s resolution is one hour and the timestamps are cycled per one week, so two check-ins which are 24*7=168 hours apart will have the same timestamps. The only learned information about the POIs is their latitudes and longitudes. These geographic coordinates enable the calculation of distance between two POIs, and the calculation is done by using haversine formula. STAN does not learn any information about the users other than their past trajectories.

We use the dataset which is used in STAN\textquotesingle s paper: the New York subset of Gowalla full dataset. The dataset has 1083 users and 5135 Points of Interest. There is only one trajectory per user.

\begin{figure}[h!]
    \centering
    \includegraphics[scale=1]{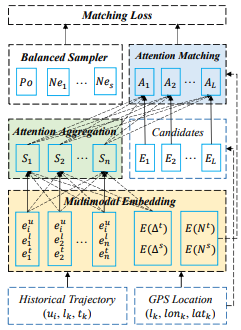}
    \caption{The schema of STAN~\cite{luo2021stan}}
    \label{fig:stan_schema}
\end{figure}

We show the schema of STAN in Figure \ref{fig:stan_schema}. The ``Multimodal Embedding'' block produces the embedding of user behavior. Thus, two users who have similar behaviors also have their embeddings close to each other. Because the user\textquotesingle s behavior comes from his/her past trajectory, the distance between the embeddings of two users represents the dissimilarity between their past trajectories. The ``Attention Matching'' block contains the information on which timesteps of the current user\textquotesingle s input trajectory are the most influential on the output. The final output is the recommended POI\textquotesingle s ID. The ``Candidates'' block contains the candidates of the POI output. STAN\textquotesingle s final layers work in the same way of many other neural network \textit{n}-class classifiers: there are \textit{n} candidates with their respective score and the final output is the candidate with the highest score. 

Collaborative filtering is a recommendation system\textquotesingle s paradigm which bases the result on the data of similar other users. Intuitively, if two users demonstrate similar behaviors in the past, their preferences should be similar as well. Another major paradigm is content-based filtering which works by learning the properties of the users\textquotesingle s choices. In the next Point of Interest (POI) recommendation system, one possible property is the POI\textquotesingle s type (e.g., restaurant, museum, night club). STAN does not use any POI property, except its latitude and longitude (i.e., the geographic location). On the other hand, STAN takes into account the user ID and learns the user behavior embedding, it means that STAN uses the collaborative filtering paradigm.

Explainability can be either global or local. Local explanations means explaining a specific prediction, so the explanation differs from one prediction to another. \cite{zhang2020explainable} classifies recommendation system explanations into ``by relevant user or item'', ``by the features of the user or the item'', ``by the existing opinions or reviews'', ``by text'', ``by the visual'', ``by the social relationship with the user'', and ``by word cluster''. Because STAN takes into account neither the features of the user/item nor opinion/review nor textual information nor visual information nor social relationship nor word cluster, therefore the only possible explanation for STAN is ``by relevant user or item''. 

\section{Related Works}
There have been works on techniques to explain the feature importance in machine learning. Shapley\textquotesingle s values~\cite{strumbelj2010efficient} assign a scalar value to quantify the feature importance. It works by permutating certain features and observing the output difference. Intuitively, if a feature is important, replacing the values of this feature will induce a significant output change. However, these permutation and observation operations are expensive, so in practice, approximation and sampling are necessary. Different extensions of Shapley\textquotesingle s values do the approximation and sampling differently. Some of the extensions of Shapley\textquotesingle s values are KernelSHAP~\cite{lundberg2017unified,covert2021improving}, FastSHAP~\cite{jethani2021fastshap}, and TreeSHAP~\cite{lundberg2020local}. Shapley\textquotesingle s values can do both global and local explanation. LIME~\cite{ribeiro2016should} is another method to explain a black box machine learning model. Fundamentally, LIME works by using a simpler but interpretable model (e.g., linear regression) to do local approximation. LIME can do local explanation only. However, when there are several features which correlate one another, there will be many correct approximations (i.e, different learned models which are all correct), so there will be many correct explanations too. For example, if $f(x_1 , x_2, x_3) = 3 x_1 + 5 x_2 + 7 x_3 + \epsilon $ while $x_1 = 2 x_3 + \epsilon$, all these approximations (or learned models) are correct: $\hat{f}(x_1 , x_2 , x_3) = x_1 + 5 x_2 + 11 x_3 + \epsilon$, or $\hat{f}(x_1 , x_2 , x_3) = 5 x_1 + 5 x_2 + 3 x_3 + \epsilon$, or $\hat{f}(x_1 , x_2 , x_3) = 5 x_2 + 13 x_3 + \epsilon$. This problem of non-unique explanation is beyond our concern.

There are also techniques which are not meant for explainability, but happen to be naturally explainable. This is the case for the attention mechanism~\cite{bahdanau2015neural}. The attention mechanism was initially introduced for the text translation problem. Both the input and the output data are expressed as variable length sequence of words. The attention mechanism is essentially a weight matrix which channels the input elements to the output elements. Because it is a weight matrix, it also explains the relative importance of the input\textquotesingle s $i$th timestep on the output\textquotesingle s $j$th timestep.

There are also works to calculate the importance of the training data points. Data Shapley~\cite{ghorbani2019data} and TracIn~\cite{pruthi2020estimating} solve this problem by using techniques from feature importance problem. Intuitively, if we imagine a 2D matrix where one axis contains the features and the other axis contains the data points, and then we run a feature importance algorithm, we will get the importance of each feature. But if we transpose the matrix before running the the feature importance algorithm, we will instead get the importance of each data point. In the context of collaborative filtering, data valuation is important because the recommendation explicitly takes into account the behavior of similar users, which are basically other data points.

\section{Methods} \label{sec:methods}
STAN\textquotesingle s output is a vector of length $L$ where $L$ is the number of Points of Interest (POIs). The final output (i.e., the recommended POI) is the argmax of that vector. This is the standard way multi-class neural network classifiers represent the output. Before those $L$ candidates, there are $T$ intermediate values where $T$ is the maximum length of the input timesteps. The ``Attention Matching'' block in Figure \ref{fig:stan_schema} is a weight matrix which channels those $T$ intermediate values to $L$ candidate POIs. This is shown in Equation \ref{eq:attention_matching_matrix} where $W$ is a weight matrix of size $L \times T$, $v$ is an intermediate vector of the size $T \times 1$, and $l_{recommended}$ is the recommended POI\textquotesingle s ID. Thus, the most important timestep is the one whose value is the highest in $W_l$ (see Equation \ref{eq:attention_matching_timestep}).

\begin{equation}
l_{recommended} = \arg \max_{l \in L} W_l v
\label{eq:attention_matching_matrix}
\end{equation}

\begin{equation}
t_{important} = \arg \max_{t \in T} W_{lt}
\label{eq:attention_matching_timestep}
\end{equation}

\begin{figure}[h!]
    \centering
    \includegraphics[scale=0.2]{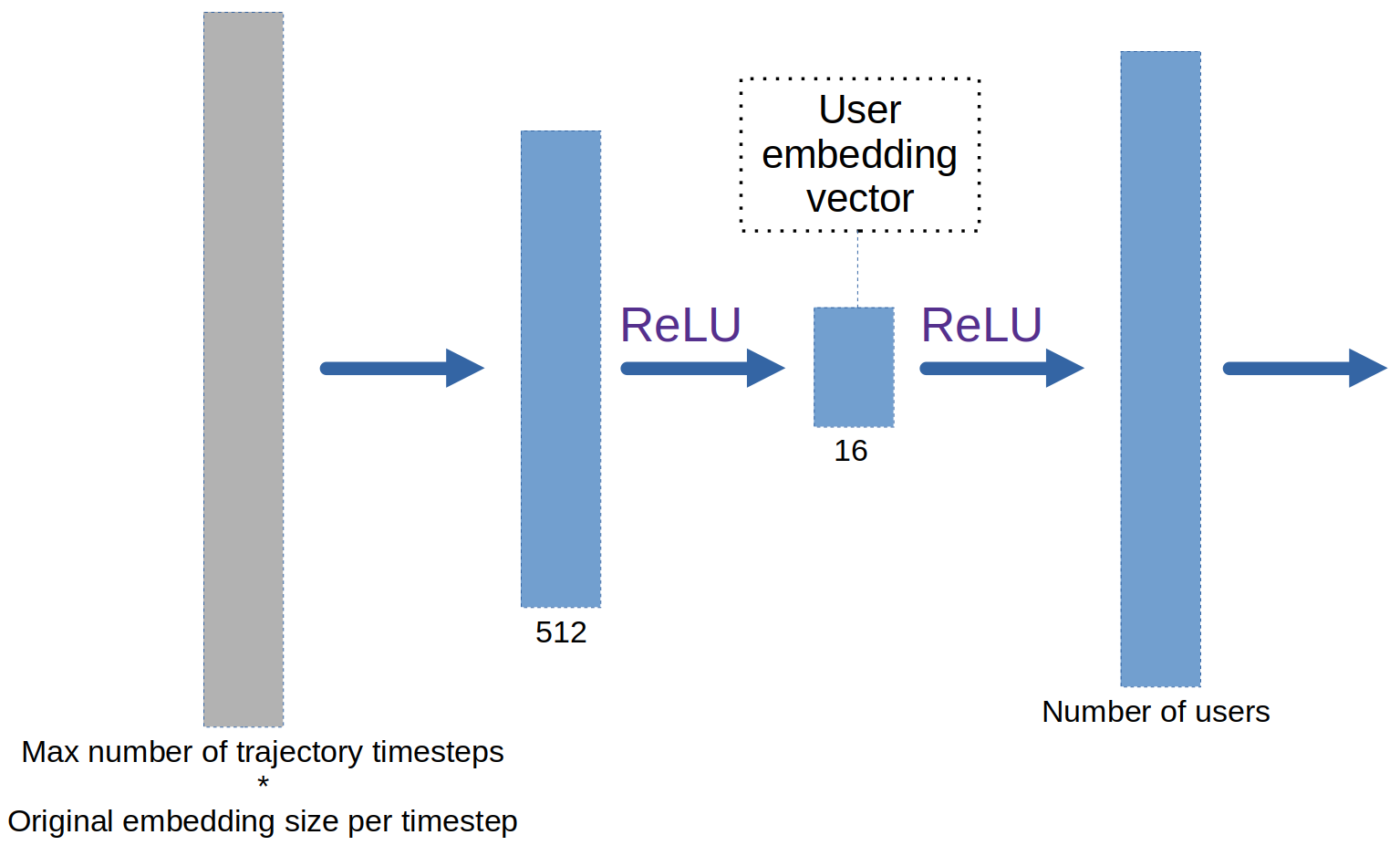}
    \caption{The schema of our compressor network which produces the 16-dimensional user embedding vector. The default original embedding size per timestep is 50 dimensions. The network learns by outputting the max value in the slot of the correct user, as is usual in multi-class neural network classifiers.}
    \label{fig:compressor_network}
\end{figure}

The user behavior embedding is shown in Figure \ref{fig:stan_schema} in the ``Multimodal Embedding'' module as $e^u$. The embeddding size is configurable, but default setting is 50 dimensions per input timestep. Because the original user embedding is for each input timestep, we create a small neural network to ``compress'' all of them into a vector of 16 dimensions per user. The network has hidden layers of 512 and 16 neurons and use ReLU activation function. The network learns by outputting the max value in the slot of the correct user (i.e., the argmax), as is usual in multi-class neural network classifiers. We show the network schema in Figure \ref{fig:compressor_network}. Onward we will refer to this network as \textbf{compressor network}. Having a representation learning network to represent a user enables us compare different users easily. Because this embedding is learned from the users\textquotesingle s past visits, similar users should have their corresponding 16-dimensional embedding vector close to each other, and similar users should also be within the arg-top-k of the output. This allows us to know which other users whose data greatly influences the recommendation for the current user, which is the essence of collaborative filtering.

\section{Experiment}
We use the standard STAN dataset (as is described in Section \ref{sec:background}). We also use STAN\textquotesingle s standard code and standard configuration to train the model.

In (\textbf{Experiment 1}), we try to find the two most important timesteps of each trajectory. Note that there is only one trajectory per user. We use the output of the ``Attention Matching'' block in Figure \ref{fig:stan_schema}. We follow the method which we describe in Section \ref{sec:methods}, Formulae \ref{eq:attention_matching_matrix} and \ref{eq:attention_matching_timestep}. In order to verify that those two timesteps are indeed important, for each of those timesteps, we replace both the POI ID and the timestamp with random values (\textbf{Experiment 1/1} and \textbf{Experiment 1/2}). Due to the sequential nature of trajectory, the random timestamp is constrained to be between the previous and the next timesteps\textquotesingle timestamps. We do this randomization 10 times and we count how many times the output (i.e., the recommended POI) changes. Then, we do a similar experiment, but we randomly choose the timestep to be randomized (\textbf{Experiment 1/3}). Intuitively, if the allegedly important timestep is indeed important, the output should change more often than when we randomly choose the timestep. We redo this experiment one more time as a sanity check (\textbf{Experiment 1/4}). We compare the frequencies of output change by using one-way ANOVA test with the p-value threshold of 0.05. The results are shown in Tables \ref{tab:randomizing_at_timestep_mean_freq_output_change} and \ref{tab:randomizing_at_timestep_pvalue}.

\begin{table}[h!]
    \centering
    \begin{tabular}{||c c||} 
     \hline
     Experiment & Average Frequency \\
     & of Output (i.e., recommended POI) Change \\ [0.5ex] 
     \hline\hline\hline\hline
     Randomizing the most & $2.982$\\
     important timestep &  \\
     Experiment 1/1 & \\
     \hline
     Randomizing the 2nd most & $2.599$\\
     important timestep &  \\
     Experiment 1/2 & \\
     \hline 
     Randomizing random & $1.813$\\
     timestep (1st run) &  \\
     Experiment 1/3 & \\
     \hline
     Randomizing random & $1.810$\\
     timestep (2nd run) &  \\
     Experiment 1/4 & \\
     \hline
    \end{tabular}
    \caption{The average frequency of output (i.e., recommended POI) change upon replacing the POI ID and the timestamp at the selected timestep (Experiment 1). The possible values are between $0$ to $10$ because we do the replacement 10 times.}
    \label{tab:randomizing_at_timestep_mean_freq_output_change}
\end{table}

\begin{table}[h!]
     \centering
     \begin{tabular}{||c c||} 
     \hline
     Compared Sub-Experiments & p-value \\ [0.5ex]
     \hline\hline\hline\hline
     Experiments 1/1 and 1/2 & $0.010$ \\
     (1st and 2nd most important) & \\
     \hline
     Experiments 1/3 and 1/4 & $0.999$ \\
     (in both experiments, replacement at random timesteps) & \\
     \hline
     Experiments 1/1 and 1/3 & $1.846 \times 10^{-20}$ \\
     (1st most important and random) & \\
     \hline
     Experiments 1/2 and 1/3 & $1.536 \times 10^{-11}$ \\
     (1st most important and random) & \\
     \hline
    \end{tabular}
    \caption{The p-values from the ANOVA tests comparing the output change frequencies upon replacing the POI ID and the timestamp at the selected timestep (Experiment 1). p-value below 0.05 indicates that the difference is statistically significant.}
    \label{tab:randomizing_at_timestep_pvalue}
\end{table}

In the results of Experiment 1, as is shown in Table \ref{tab:randomizing_at_timestep_mean_freq_output_change}, we find that randomizing the POI ID and the timestamp at a the most important timestep (Experiment 1/1) induces more frequent output (i.e., recommended POI) change than when the randomization happens at the 2nd most important timestep ($2.982$ vs $2.599$). The change frequency is even lower when the timestep is chosen randomly in Experiments 1/3 and 1/4 ($1.813$ and $1.810$). We notice as well in Table \ref{tab:randomizing_at_timestep_pvalue} that all the differences have a p-value which is below our threshold of $0.05$, except for the difference between Experiments 1/3 and 1/4, which are essentially the same stochastic experiment done twice.

In (\textbf{Experiment 2}), for each user, we try to find the two most similar other users. This is from the arg-top-2 (other the current user him/herself) of the ``compressor network''\textquotesingle s output (see Figure \ref{fig:compressor_network}). Intuitively, given the trajectory of a certain user, if we replace the user\textquotesingle s ID with the ID of a similar user, the recommended POI will be less likely to change than if we replace it with the ID of a random user. This intuition comes from the essence of collaborative filtering: the recommendation is based on the user\textquotesingle s past behavior and the behavior of similar other users. Because there is only one trajectory for each user, we assign another user\textquotesingle s ID to this trajectory, and then we check if the recommended POI remains the same. We expect that assigning a similar user\textquotesingle s ID will induce less frequent changes on the recommended POI than assigning a random user\textquotesingle s ID. In \textbf{Experiment 2/1}, we replace the user\textquotesingle s ID with the ID of the most similar user and we calculate the proportion of the recommendations which remain the same. In \textbf{Experiment 2/2}, we replace the user ID with the ID of the 2nd most similar user. In \textbf{Experiment 2/3}, we still do a similar procedure, but we randomly choose the replacement user (we do this 30 times). As a sanity check, we redo the procedure of the replacement with random users (\textbf{Experiment 2/4}). We compare the frequencies of unchanged output/recommendation by using Student\textquotesingle s t-test with the p-value threshold of 0.05. The results are shown in Tables \ref{tab:assigning_different_user_proportion_recommended_poi_unchanged} and \ref{tab:assigning_different_user_proportion_recommended_poi_unchanged_pvalue}.

\begin{table}[h!]
    \centering
    \begin{tabular}{||c c||} 
     \hline
     Experiment & Proportion of Recommended \\
     & POI Being Unchanged \\ [0.5ex] 
     \hline\hline\hline\hline
     Assigning The Most & $3.693\%$\\
     Similar User ID &  \\
     Experiment 2/1 & \\
     \hline
     Assigning The 2nd Most & $3.970\%$\\
     Similar User ID &  \\
     Experiment 2/2 & \\
     \hline 
     Assigning Random & $2.684\%$\\
     User IDs 30x (1st run) &  \\
     Experiment 2/3 & \\
     \hline
     Assigning Random & $2.604\%$\\
     User IDs 30x (2nd run) &  \\
     Experiment 2/4 & \\
     \hline
    \end{tabular}
    \caption{The proportion of unchanged outputs / recommended POIs upon the assignment of another user IDs (Experiment 2)}
    \label{tab:assigning_different_user_proportion_recommended_poi_unchanged}
\end{table}

\begin{table}[h!]
    \centering
    \begin{tabular}{||c c||} 
     \hline
     Compared Sub-Experiments & p-value \\ [0.5ex]
     \hline\hline\hline\hline
     Experiments 2/1 and 2/3 & $1.257 \times 10^{-11}$ \\
     (the most similar and random) & \\
     \hline
     Experiments 2/2 and 2/3 & $3.628 \times 10^{-14}$ \\
     (the most 2nd similar and random) & \\
     \hline
     Experiments 2/3 and 2/4 & $0.558$ \\
     (in both experiments, assigning random User IDs) & \\
     \hline
    \end{tabular}
    \caption{The p-values from the Student\textquotesingle s t-tests of the comparison of the proportion of unchanged outputs / recommended POIs upon the assignment of another user ID (Experiment 2). p-value below 0.05 indicates that the difference is statistically significant.}
    \label{tab:assigning_different_user_proportion_recommended_poi_unchanged_pvalue}
\end{table}

In the results of Experiment 2, as are shown in Table \ref{tab:assigning_different_user_proportion_recommended_poi_unchanged}, we see that that replacing the user IDs with similar users\textquotesingle IDs (Experiments 2/1 and 2/2) induces less frequent output / recommendation changes than when the replacement user ID is chosen randomly (Experiments 2/3 and 2/4), and the p-values are below 0.05 (see Table \ref{tab:assigning_different_user_proportion_recommended_poi_unchanged_pvalue}, ``Experiments 2/1 and 2/3'', and ``Experiments 2/2 and 2/3''). We also see in Table \ref{tab:assigning_different_user_proportion_recommended_poi_unchanged_pvalue} that ``Experiments 2/3 and 2/4'', which are basically the same stochastic experiment done twice, have a p-value of more than the threshold 0.05.

In \textbf{Experiment 3}, we try to figure out what makes users to be similar. We might say that two users are similar if they have visited the same POIs. However, in the case of STAN, the geographical distance between the POIs are explicitly considered. Specifically, STAN calculates the geographical distance by using haversine formula based on the latitude and longitude of the POIs. Thus, we can hypothesize that the past POIs of two similar users are likely to be near to each other. The dataset also includes the visit timestamps (at the granularity of one hour, cycled every one week). So, it is also a reasonable hypothesis that the visit timestamps of two similar users are close to each other. In \textbf{Experiment 3/1/1}, we measure the average geographic/haversine distance between all the POIs which a user has visited and all the POIs which have been visited by the most similar other user (see Formula \ref{eq:average_geographic_distance}, $P_i$ is the set of POIs visited by user $i$, $P_j$ is the set of POIs visited by user $j$), and then we compare the aforementioned distance against the average distance to 10 random users. The comparison is performed by using Student\textquotesingle s t-test with the p-value threshold of 0.05. A p-value below this threshold indicates a statistically significant difference. Based on this p-value threshold, we measure the proportion of users whose distance to the most similar other user is statistically-significantly closer than the distance to random users. This comparison is between one deterministic value (the POI distance to the most similar user) and 10 stochastic values (because we choose 10 random users as comparators). In \textbf{Experiment 3/1/2}, instead of measuring the average geographic/haversine distance, we measure the average timestamp difference (see Formula \ref{eq:average_timestamp_difference}, $T_i$ is the set of timestamps of user $i$, $T_j$ is the set of timestamps of user $j$). Note that the timestamps are already rounded to the granularity of one hour and is cycled everyone week. In \textbf{Experiment 3/2/1}, we measure the average POI distances again, but this time we only consider the 10 closest POI pairs (for each user pairing). In \textbf{Experiment 3/2/2}, we also take the 10 closest pairs for each user pairing, but the closeness is on the timestamp difference instead of the geographic distance. We show the results in Table \ref{tab:measuring_distance_to_similar_users}.

\begin{equation}
distance_{geographic, average} = \frac{1}{|P_i| |P_j|}\sum_{p_i \in P_i}\sum_{p_j \in P_j} haversine(p_i, p_j)
\label{eq:average_geographic_distance}
\end{equation}

\begin{equation}
difference_{timestamp, average} = \frac{1}{|T_i| |T_j|}\sum_{t_i \in T_i}\sum_{t_j \in T_j} |t_i - t_j|
\label{eq:average_timestamp_difference}
\end{equation}

\begin{table}[h!]
    \centering
    \begin{tabular}{||c c||} 
     \hline
     Experiment & Proportion of Statistically-Significant Differences \\
     & (The Most Similar User and \\
     & The 2nd Most Similar User) \\ [0.5ex] 
     \hline\hline\hline\hline
     POI Distances & $34.257\%$\\
     (all) & $35.642\%$ \\
     Experiment 3/1/1 & \\
     \hline
     Timestamp Difference & $30.286\%$\\
     (all) & $29.086\%$ \\
     Experiment 3/1/2 & \\
     \hline 
     POI Distances & $25.762\%$\\
     (10 closest pairs) & $26.038\%$ \\
     Experiment 3/2/1 & \\
     \hline
     Timestamp Difference & $32.041\%$\\
     (10 closest pairs) &  $31.579\%$\\
     Experiment 3/2/2 & \\
     \hline 
    \end{tabular}
    \caption{The proportion of the occurrences when the distances to the most similar user and to the 2nd most similar user are statistically-significantly closer than the distances to 10 random users as measured by Student\textquotesingle s t-test with the p-value threshold of 0.05 (Experiment 3). The difference is between one deterministic value (the POI-distance / timestamp-difference to the most / 2nd-most similar user) and 10 stochastic values (because there are 10 randomly chosen users).}
    \label{tab:measuring_distance_to_similar_users}
\end{table}

We see the results of Experiment 3 in Table \ref{tab:measuring_distance_to_similar_users} show that whether we use POI distances or timestamp differences, whether we take all pairs or only the 10 closest pairs, only around 30\% of the cases show a statistically significant difference between the distance/difference to the most / 2nd-most similar user and to 10 randomly chosen users. This led us to Experiment 4 where we use a synthetic ``dumbed-down'' data where the user similarities are deliberately made trivial, so the most and the 2nd-most similar users should be predictable.

In \textbf{Experiment 4}, we construct a synthetic data with predictable similarity relationship. Specifically, out of the 1083 original users (whose user ID ranges from 1 to 1083, inclusive of both numbers), we replace those whose user ID is between 101 and 300 (inclusive of both numbers) to be the almost-clones of those whose user ID is between 1 and 100 (inclusive of both numbers). For each $user_{i}$ where $1 \le i \le 100$, we replace $user_{i+100}$ such that the past visits of users $i$ and $i+100$ have the same length, the same timestamps, while the POIs differ at one timestep only. We choose the replacement POI to be the geographically nearest one from the original POI. Other than that, we also replace $user_{i+200}$ such that the past visits of users $i$ and $i+200$ have the same length, the same POIs, while the timestamps differ at one timestep only. We choose the replacement timestamp to differ by one hour only from original timestep. Note that the timestamps are already rounded to the granularity of one hour and is cycled every one week, thus one hour is the smallest possible difference. The result of this process is that the users $i+100$ and $i+200$ are almost-clones of $user_i$, which means they should be the most and the 2nd-most similar users of user $i$. We then re-train the STAN model by using this new dataset. With the new trained model, for all $user_{i}$ where $1 \le i \le 100$, we count how many times the most similar \textbf{OR} the 2nd most similar user is either $user_{i+100}$ \textbf{OR} $user_{i+200}$. Because there are 1083 users, the probability that at least one of two randomly chosen other users is either the most similar or the 2nd most similar users is $\frac{4}{1083-1} \approx 0.0037$. We show the result of this experiment in Table \ref{tab:counting_similar_users}.

\begin{table}[h]
    \centering
    \begin{tabular}{||c c||} 
     \hline
     Note & Count \\ [0.5ex] 
     \hline\hline\hline\hline
     Experiment 4 & $1$\\
     \hline
    \end{tabular}
    \caption{For the 100 $user_i$ in the synthetic data of Experiment 4 where $1 \le i \le 100$, we count how many times the most similar OR the 2nd most similar user is either $user_{i+100}$ OR $user_{i+200}$ (who are both the almost-clones of $user_i$). At chance, the expected count would be $100 \times \frac{4}{1083-1} \approx 0.370$}
    \label{tab:counting_similar_users}
\end{table}

The result of Experiment 4 in Table \ref{tab:counting_similar_users} shows that the count is only barely higher than chance ($1$ against $0.370$), even though those 100 users have 200 known almost-clones.

\section{Conclusion}

In Experiment 1, as is shown in Table \ref{tab:randomizing_at_timestep_mean_freq_output_change}, we find that randomizing the POI ID and the timestamp at a carefully chosen timestep (i.e., Experiments 1/1 and 1/2) changes the output / recommendation more frequently than when the timestep is randomly selected (i.e., Experiments 1/3 and 1/4). The p-values as are shown in Table \ref{tab:randomizing_at_timestep_pvalue} also show that the differences are statistically significant. This demonstrates that STAN\textquotesingle s ``Attention Matching'' correctly reveals which input timesteps are important.

The results of Experiment 2 which we show in Tables \ref{tab:assigning_different_user_proportion_recommended_poi_unchanged} and \ref{tab:assigning_different_user_proportion_recommended_poi_unchanged_pvalue} show that replacing the user IDs with similar users\textquotesingle IDs causes less frequent recommendation changes than when the replacement is a randomly chosen user ID. The differences are also statistically significant. This shows that the user behavior embedding learned by STAN\textquotesingle s ``Multimodal Embedding'' module is correctly taken into account by the downstream parts of STAN. Effectively, it means that when two users are similar in terms of the embedding, their final outputs (i.e., the recommendation) are more likely to be the same as well.

On the other hand, the results of Experiment 3 which we show in Table \ref{tab:measuring_distance_to_similar_users} suggest that neither POI geographic distances nor timestamp differences explains the supposed similarity between the users. In only around 30\% of the cases we see a statistically significant difference between the most or the 2nd-most similar user and random users. This led us to Experiment 4 where we use a synthetic ``dumbed-down'' data where the user similarities are trivial, and thus are easier to predict.

In Experiment 4 where we use a synthetic dataset where for 100 known users, each of them have two almost-clones. However, as is shown in Table \ref{tab:counting_similar_users}, the count is only barely higher than chance ($1$ against $0.370$). Besides that, the count is also underwhelming because those 100 known users have total 200 known almost-clones. This suggests that STAN\textquotesingle s user embedding module, at least in the default setting, is not trained very well.

We have demonstrated that we can make STAN explainable, essentially just by looking into its code base. We do not use any stand-alone complex nor expensive explainability technique for this. We demonstrate as well that given the context of the problem which the recommendation system tries to solve, explainability is useful to ``debug'' the model. For example, in a technique based on collaborative filtering, it is useful to ask if the allegedly similar other users are indeed similar to the current user.

In this work, we observe the importance of understanding of the problem\textquotesingle s context. In the case of next POI recommendation system, the past trajectory of other users are not really features from the point of view of machine learning, yet this is an important piece of information if the recommendation system uses collaborative filtering. Indeed, we find in Experiments 3 and 4 that the user similarity notion of our STAN model is questionable even though this is an important component of collaborative filtering. Stand-alone (post-hoc) feature importance techniques are unlikely to reveal this problem.

\section*{Acknowledgement}
We thank Jiaye Song and Pratik Karmakar whose prior work inspired us to pursue this research project.

\bibliographystyle{unsrt}  
\bibliography{references}  

\end{document}